\documentclass[final]{svjour3}
\usepackage{graphicx}
\usepackage{rotating}
\usepackage{amssymb}
\usepackage{mathptmx}
\usepackage{xcolor}
\usepackage[numbers]{natbib}
\usepackage{subfigure}
\usepackage{color, soul}
\makeatletter
\journalname{Journal of Low Temperature Physics}

\bibpunct{[}{]}{,}{n}{}{,}

\begin{document}

\newcommand{\hdblarrow}{H\makebox[0.9ex][l]{$\downdownarrows$}-}
\title{Modeling a Three-Stage SQUID System in Space with
the First Micro-X Sounding Rocket Flight}

\author{
Adams,~J.S. \and 
Bandler,~S.R. \and 
Bastidon,~N. \and 
Eckart,~M.E. \and 
Figueroa-Feliciano,~E. \and 
Fuhrman,~J.* \and 
Goldfinger,~D.C. \and 
Hubbard,~A.J.F. \and 
Jardin,~D. \and 
Kelley,~R.L. \and 
Kilbourne,~C.A. \and 
Manzagol-Harwood,~R.E. \and 
McCammon,~D. \and 
Okajima,~T. \and
Porter,~F.S. \and 
Reintsema,~C.D. \and 
Smith,~S.J.
}

\institute{*Corresponding Author\\
\email{joshuafuhrman2023@u.northwestern.edu}}

\maketitle

\begin{abstract}

The Micro-X sounding rocket is a NASA funded X-ray telescope payload that completed its first flight on July 22, 2018. This event marked the first operation of Transition Edge Sensors (TESs) and their SQUID-based multiplexing readout system in space. Unfortunately, due to an ACS pointing failure, the rocket was spinning during its five minute observation period and no scientific data was collected. However, data collected from the internal calibration source marked a partial success for the payload and offers a unique opportunity to study the response of TESs and SQUIDs in space. Of particular interest is the magnetic field response of the NIST MUX06a SQUID readout system to tumbling through Earth’s magnetic field. We present a model to explain the baseline response of the SQUIDs, which lead to a subset of pixels failing to "lock" for the full observational period. Future flights of the Micro-X rocket will include the NIST MUX18b SQUID system with dramatically reduced magnetic susceptibility.


\end{abstract}

\section{Introduction}

The first flight of the Micro-X sounding rocket \cite{Wikus:2010} occurred in July 2018, marking the first time Transition Edge Sensors (TES) were operated in space~\cite{Adams:2019, Adams:2021}. A failure of the Attitude Control System (ACS) caused Micro-X to tumble in space for its full five minute observation. While no scientific data was obtained, the rotation of the rocket through Earth’s magnetic field offers a unique opportunity to study how TESs and their SQUID-based readout systems respond in space. We present a model to describe the system’s response to external magnetic fields based on the flown NIST MUX06a SQUIDs, a three-stage time-division multiplexing (TDM) system \cite{Stiehl:2011, Korte:2003}.

\begin{figure}[htbp]
\begin{center}
\includegraphics[width=0.45\linewidth,keepaspectratio]{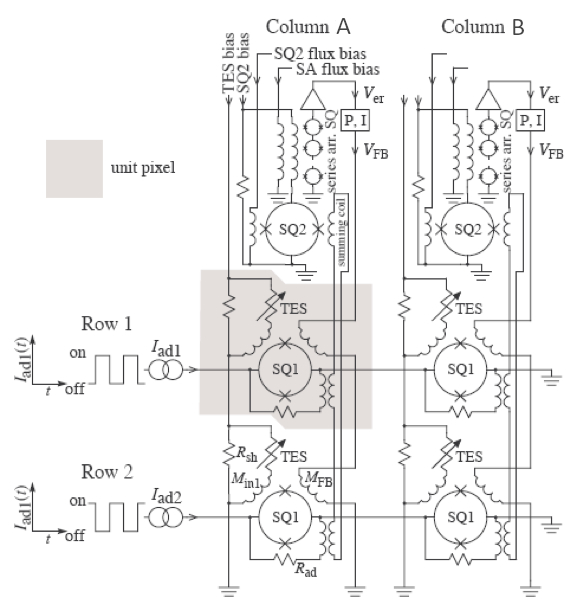}
\caption{Circuit Diagram for the Micro-X detectors and MUX06a SQUID readout. Each TES pixel is periodically read out by a SQ1, which is amplified by a SQ2 and SA unique to a column of 16 pixels. The four column-specific tuning parameters are the TES bias current, the SQ2 bias current, the SQ2 flux bias, and the SA flux bias. The SQ2 flux bias changes where the output of SQ1 maps onto the input of the SQ2 V-$\Phi$. The superconducting summing coil couples individual SQ1s to the column-specific SQ2, and is the primary point of magnetic susceptibility in the circuit. External magnetic fields that couple to the system act as an additional SQ2 flux bias. Republished from~\cite{Wikus:2009}.}
\label{fig:squid_circuit}
\end{center}
\end{figure}

The Micro-X TES array~\cite{Eckart:2009} is read out through a three-stage SQUID system in a flux-locked loop, shown in Fig.~\ref{fig:squid_circuit}. A characteristic property of SQUIDs is their periodic voltage response (V) to magnetic flux ($\Phi$), or their V-$\Phi$ relation, with a period equal to the magnetic flux quantum $\Phi_0$. The 128 TES pixels are divided into two independent science chains (X and Y), arranged in four columns (A--D) of 16 pixels each. Each pixel is read out by a first stage SQUID (SQ1) unique to the pixel, and then periodically read out by a second stage SQUID (SQ2) and a SQUID array (SA), both unique to the column. This SQUID system has several properties that the model aims to reflect. These include: asymmetric V-$\Phi$ relationships that have a steep, high-gain side, and a more gradual, low-gain side; SQ1s that are strongly coupled to the SQ2 input such that the high-gain domain of the SQ2 is fully covered by the output of SQ1; SAs that have an approximately linear response over the output range of the SQ2s; and a magnetic response that is dominated by the effective area of the SQ2 summing coil used to couple the SQ1s to the SQ2 in each column~\cite{Stiehl:2011}. Additionally, the system is tuned to utilize the high-gain side of the SQ2s.

\section{Three-Stage SQUID Model}
\label{sec:model}

\begin{figure}[htbp]
\begin{center}
\includegraphics[width=0.65\linewidth,keepaspectratio]{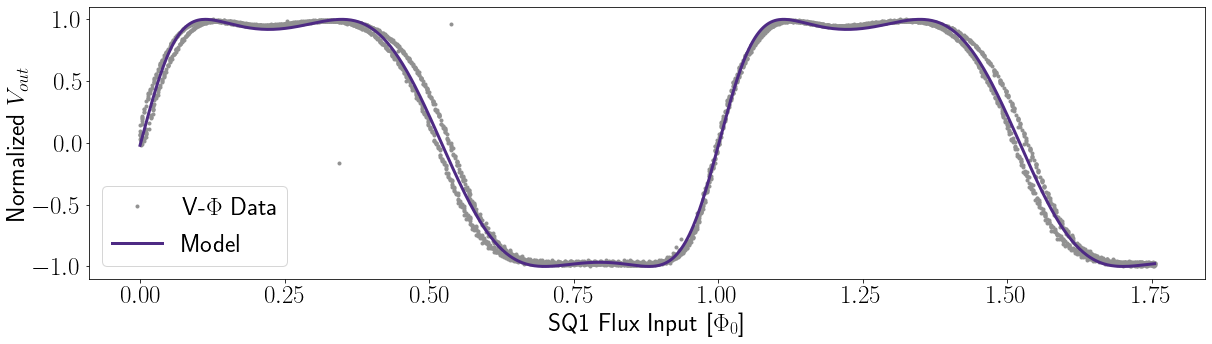}
\caption{Model fit to a composite V-$\Phi$ for YC01, a pixel characteristic of the array, tuned on the high-gain size of SQ2. Fit parameters $n_1$=2, $n_2$=7, $scale$=1.2. (Color figure online.)}
\label{fig:squid_fit}
\end{center}
\end{figure}

To reflect the asymmetric V-$\Phi$ response we will model each SQUID stage as a truncated sawtooth wave summation, Eq.~(\ref{eqn:sawtooth}). Increasing the number of terms in the summation, n, increases how skewed the V-$\Phi$ will be. The input to each SQUID stage will be modeled in units of $\Phi_0$, and the output voltage will be normalized between $\pm1$. The three major model parameters are the number of SQ1 and SQ2 sawtooth terms, $n_1$ and $n_2$, and the $scale$ that maps the output of SQ1 onto the input of SQ2. The model parameters are fit to the real SQUIDs using composite V-$\Phi$ data. Typical values and an example of data fit to the model output are shown in Fig.~\ref{fig:squid_fit}. 

\begin{equation}
	\sum_{k=1}^{n} \frac{{2n \choose n-k}}{{2n \choose n}} \frac{\sin{kx}}{k}
	\label{eqn:sawtooth}
\end{equation}

\begin{figure}[htbp]
\begin{center}
\includegraphics[width=0.55\linewidth,keepaspectratio]{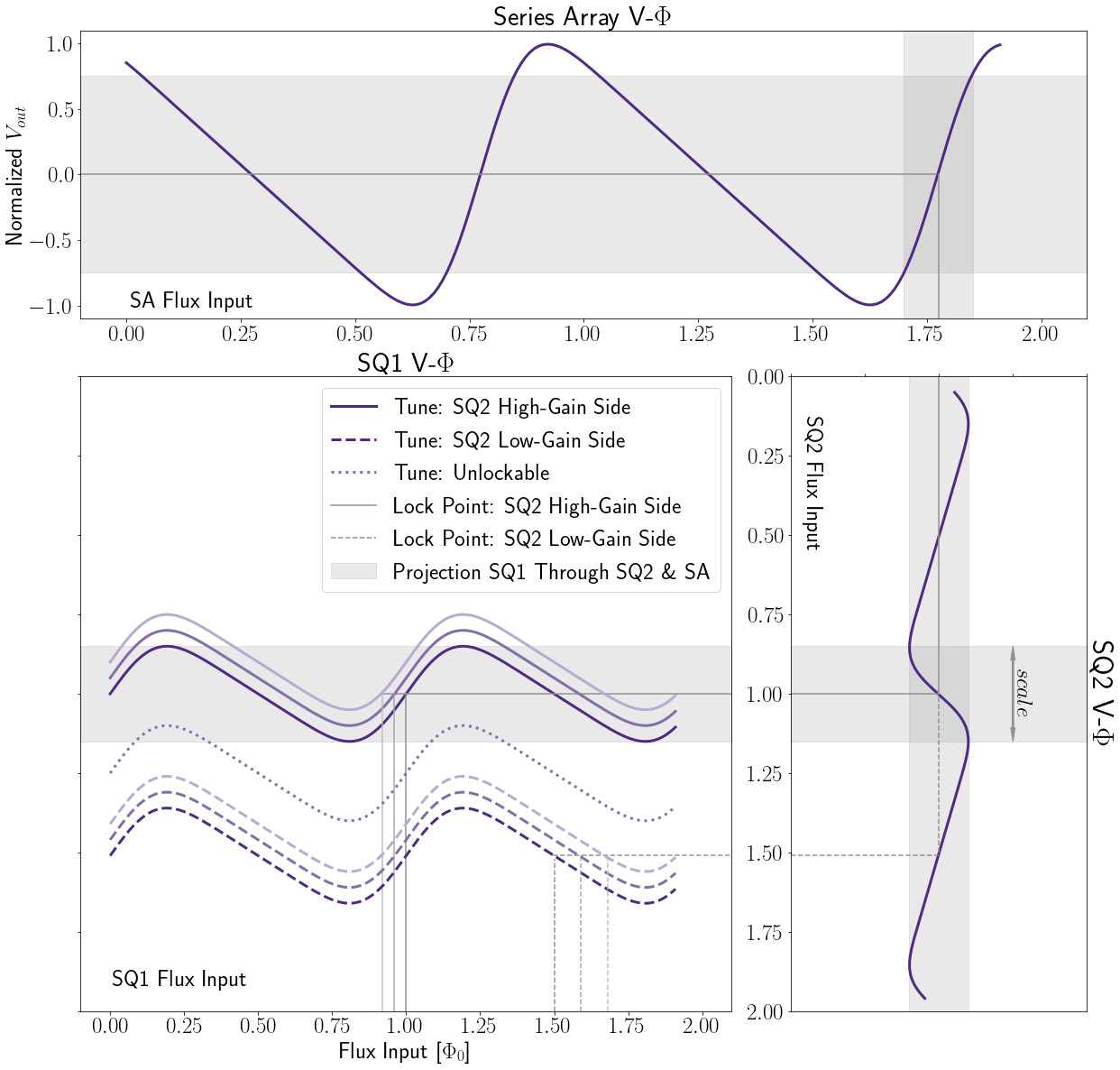}
\caption{Model relationship between the three SQUID stages. SQ1 V-$\Phi$s tuned on the high-gain side of the SQ2 are shown in \textit{solid purple} and their corresponding locked baseline values in \textit{solid grey}. Those tuned on the low-gain side of the SQ2 are shown in \textit{dashed purple} and \textit{dashed grey}. A SQ1 without a stable lock point is shown in \textit{dotted purple}. The projection of a properly tuned SQ1 through the SQ2 and SA is highlighted in \textit{grey}. The model \textit{scale} parameter is shown where the SQ1 projects onto the SQ2 V-$\Phi$ \textit{bottom right}. (Color figure online.)}
\label{fig:3stage_model}
\end{center}
\end{figure}

The structure of the model is shown in Fig.~\ref{fig:3stage_model}. For a given flux input the SQUID response can be traced through a SQ1 V-$\Phi$ (\textit{bottom left}), the SQ2 \mbox{V-$\Phi$} (\textit{bottom right}), and the SA \mbox{V-$\Phi$} (\textit{top}) to produce the normalized voltage output. When operating in a flux-locked loop, the voltage output is used by the PI controller to adjust the flux input, which establishes a locked baseline value. By adjusting the relationship between the modeled SQUID stages we can simulate changing the SQUID tuning parameters described in Fig.~\ref{fig:squid_circuit}. The most important tuning parameter is the SQ2 flux bias, which determines where the output of SQ1 maps onto the input of SQ2. The large effective area of the SQ2 summing coil dominates the coupling of the SQUIDs to external magnetic fields. This means the effect of external magnetic fields on the SQUIDs can be modeled as a change in the SQ2 flux bias. DC offsets to the SQ2 flux input produced in this way effectively shift the SQ1 V-$\Phi$ up or down, represented by the various SQ1 V-$\Phi$s in Fig.~\ref{fig:3stage_model}. If the change in the SQ2 flux bias is sufficiently large, then there may not exist a stable SQ1 flux input for the flux-locked loop, as shown by the dotted SQ1 curve. When this happens the SQUIDs ``unlock" and cannot read out that TES. If the SQ2 flux bias continues to change, then the system may relock on  the low-gain side of the SQ2. When this happens, the locked baseline will also jump to the low-gain side of the SQ1 to preserve the PI relationship between the SQ1 flux input and the normalized voltage output.

\begin{figure}[htbp]
\begin{center}
\begin{subfigure}
\centering
\includegraphics[width=0.4\linewidth,keepaspectratio]{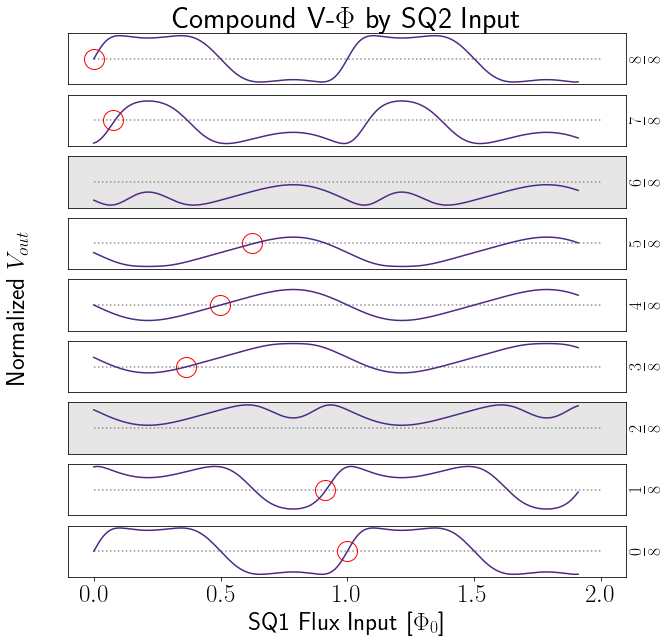}
\end{subfigure}
\begin{subfigure}
\centering
\includegraphics[width=0.4\linewidth,keepaspectratio]{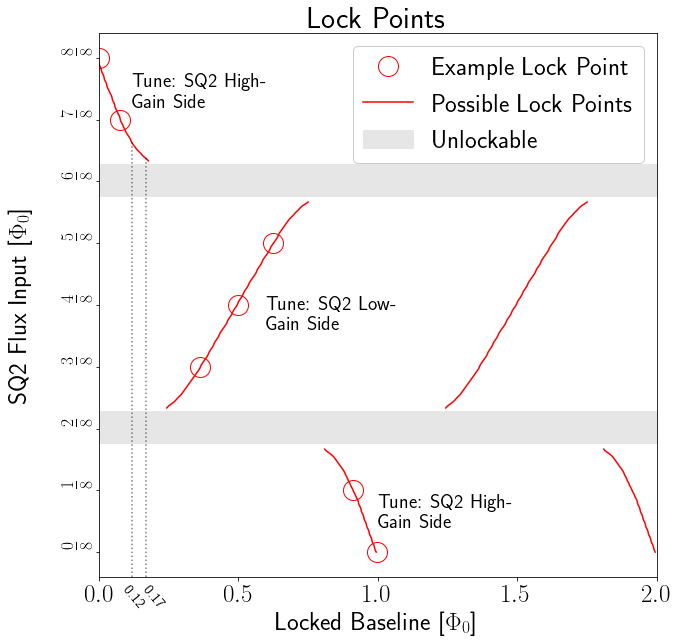}
\end{subfigure}
\caption{(a), \textit{left}: Modeled composite V-$\Phi$ curves and locked baseline values at various SQ2 flux inputs. (b), \textit{Right}: Parameter space showing the relationship between SQ2 flux input and locked baseline values. Values highlighted in \textit{grey} have no stable lock points. Lock points from the example V-$\Phi$ curves (\textit{left}) marked in \textit{red circles}. (Color figure online.)}
\label{fig:param_space}
\end{center}
\end{figure}

Using this model, a composite V-$\Phi$ can be calculated for different SQ2 flux biases, along with any possible locked baseline values. Example composite V-$\Phi$ relationships are detailed in Fig.~\ref{fig:param_space}a, and the corresponding phase space in Fig.~\ref{fig:param_space}b. For a nominally tuned SQUID, as the SQ2 flux bias changes from zero the baseline follows the red curve through phase space with an initially linear response of 1.4e-2~$\Phi_{0}$~per~mG through the SQ2 summing loop, assuming a SQ2 effective area of 468.5~$\mu m^2$~\cite{Stiehl:2011}. This response increases after a 0.12~$\Phi_0$ baseline shift, and unlocks in the highlighted region after a 0.17~$\Phi_0$ baseline shift. Similar behaviour is expected when the SQUIDs relock on the low-gain side of the SQ2, with a few notable differences. The baseline will relock about 0.5~$\Phi_0$ away from where it unlocked as the PI loop forces the lock point to the low-gain side of SQ1. Additionally, the locked baseline response is greater and in the opposite direction on the SQ2 low-gain side as it is on the high-gain side. This is shown clearly in the projected locked baseline values in Fig.~\ref{fig:3stage_model}, where equal steps in SQ2 flux input produce small decreases in baseline on the high-gain side of the SQ2, and larger increases in baseline on the low-gain side. The linear baseline response on the SQ2 low-gain side is 2.3e-2~$\Phi_0$~per~mG throgh the SQ2 summing loop.

\section{Observed Flight Effects}

This model can explain many of the anomalous system effects observed in flight due to the instrument tumbling in Earth's magnetic field.

\subsection{SQUID Unlocking}

\begin{figure}[htbp]
\begin{center}
\includegraphics[width=0.65\linewidth,keepaspectratio]{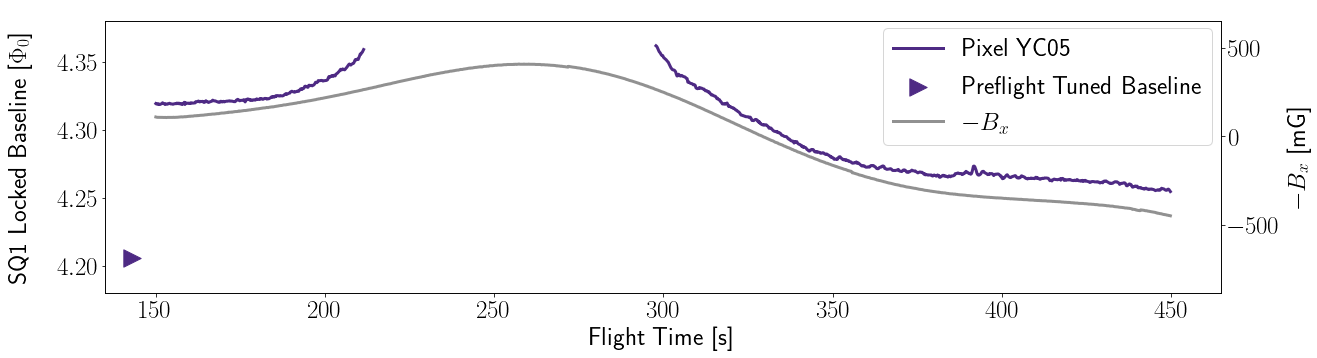}
\caption{Example locked baseline response during observation (\textit{left axis}) and magnetic field in the negative rocket x-direction (\textit{right axis}). Time is referenced from launch. This SQUID was unlocked between 212~s and 298~s. The \textit{purple triangle} marks the pixel's baseline when it was well tuned before flight. (Color figure online.)}
\label{fig:obs_unlocking}
\end{center}
\end{figure}

During the five-minute observation period many of the SQUIDs unlocked for approximately two minutes. Additionally, the locked SQ1 baseline appears to track with the external magnetic field in the negative rocket x-direction. The baseline correlation with magnetic fields in this direction specifically is due to the intricacies of the detector's superconducting magnetic shielding, and will be discussed in more detail in an upcoming instrument paper reviewing the  first flight of Micro-X. An example of this response is shown in Fig.~\ref{fig:obs_unlocking} and is explained by several features of the model. The marked pre-flight baseline value should be considered well tuned on the high-gain side of the SQ2, and any changes in the baseline due to Earth's magnetic field should be referenced from this point. The difference between the pre-flight baseline and the baseline at the start of observation is due to the different orientation of the rocket in Earth's magnetic field. As the locked baseline drifted during observation, the SQUIDs unlocked when the correlated magnetic field value is at a maximum. Additionally, the baseline response is observed to shift more rapidly before and after the SQUIDs unlocking - a clear prediction of the modeled SQUID relationships. The baseline response is observed to increase after drifting 0.12~$\Phi_0$, and unlock after drifting 0.16~$\Phi_0$. This is consistent with the model fit to these SQUIDs, and can be seen in Fig.~\ref{fig:param_space}b.

\subsection{Baseline Oscillation in Powered Flight}

\begin{figure}[htbp]
\begin{center}
\includegraphics[width=0.65\linewidth,keepaspectratio]{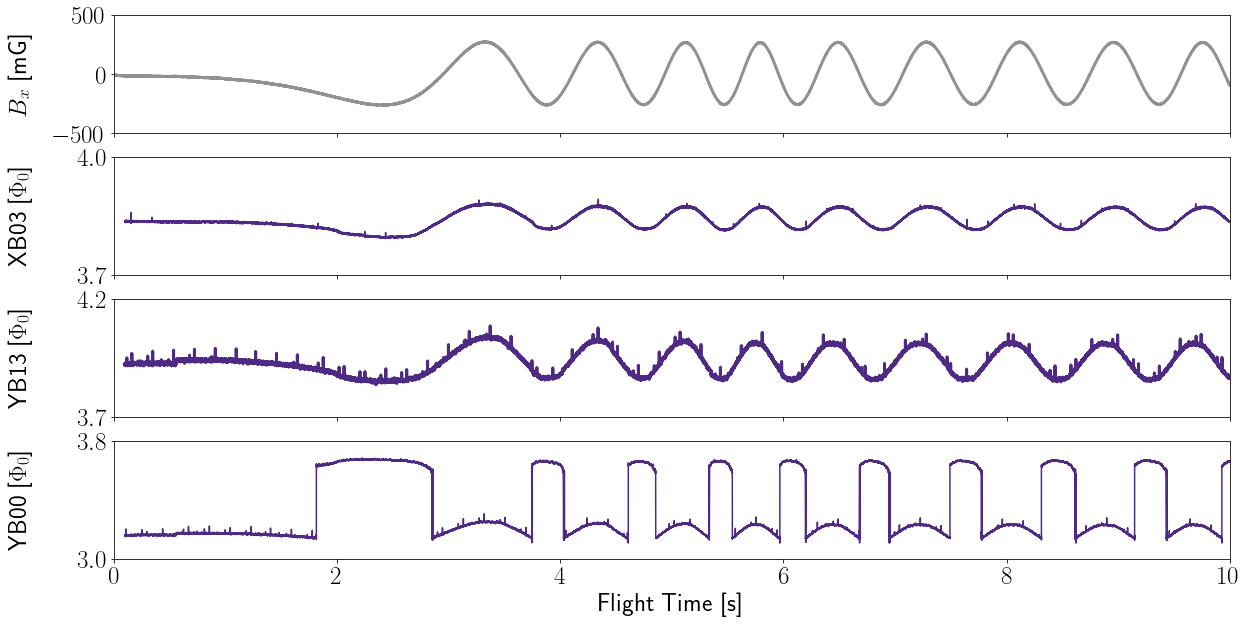}
\caption{Example locked baseline responses during powered flight (\textit{purple}) and magnetic field in the rocket x-direction (\textit{grey}). Pixel XB03 remained locked on the high-gain side of the SQ2, while pixel YB13 relocked on the low-gain side. Pixel YB00 jumps between lock points on the high-gain and low-gain sides of SQ2 as the rocket spins. Time is referenced from launch. (Color figure online.)}
\label{fig:powered_flight}
\end{center}
\end{figure}

During powered flight, the rocket spins about the thrust (z) axis at approximately 4~Hz for stability. The external magnetic field in the rocket x-direction produces an oscillation on the locked feedback baseline as shown for pixels XB03 and YB13 in Fig.~\ref{fig:powered_flight}. Note that the detectors are tuned and operate at 75~mK, but are kept at 300~mK during powered flight. Therefore, the ADR magnet current and the SQ2 flux input are arbitrary at this time. Some pixels continued to read out normally (XB03), while others either unlocked or relocked on the low-gain side of their SQ2 (YB13). The locked baseline of pixel XB03 tracks the magnetic field in the positive rocket x-direction at 1.1e-4~$\Phi_{0}$~per~mG, while pixel YB13 tracks at 3.3e-4~$\Phi_{0}$~per~mG.\footnote{In addition to the baseline response to external magnetic fields inverting between the high-gain and low-gain sides of the SQ2, the response also inverts between detector sides X and Y. This was verified via post-flight testing with a Helmholtz coil, and is likely due to the spatial layout of the MUX06a chips and bending of the field inside the superconducting shielding. This will be discussed in more detail in an upcoming instrument paper reviewing the first flight of Micro-X.} The model predicts that the baseline response should be 1.6 times greater on the SQ2 low-gain side than on the high-gain side. The observed ratio of baseline responses is therefore higher than would be expected if XB03 and YB13 were in the linear regions of the SQ2 high- and low-gain side, respectively. However, all of column XB remained well behaved during powered flight and was most likely in the linear region, while column YB was in the non-linear region as evidenced by pixel YB00 in Fig.~\ref{fig:powered_flight}. A larger ratio of responses is therefore expected.

As the SQ2 flux bias changes, if a SQ1 is sufficiently coupled to the SQ2, it might not unlock and instead jump directly between the two sides of the SQ2. In doing so, the baseline response will change in magnitude and reverse direction, and the size of the baseline jumps will be approximately 0.5~$\Phi_0$ as discussed in Section~\ref{sec:model}. Both of these features are clearly visible in the baseline of pixel YB00.

\section{Conclusions}

Modeling the relationships between the three MUX06a SQUID stages helps explain many of the behaviours observed in the first Micro-X flight. The primary source of magnetic susceptibility in the system was the superconducting summing coil that connects SQ1 to SQ2, increasing the effective area of the SQ2 stage. Treating the external magnetic field as an additional SQ2 flux input models many of the effects observed in flight: the pixels can become unlocked if the SQ1 output is shifted to a state where no stable lock point exists; before pixels unlock, the locked baseline has an increased sensitivity to the external field; and pixels that lock on the low-gain side of SQ2 have an increased and inverted baseline response.

For the upcoming second Micro-X flight in 2022 the NIST MUX06a SQUIDs have been replaced with the less magnetically susceptible NIST MUX18b SQUIDs~\cite{Reintsema:2019}. See the upcoming instrument paper for a full review of the Micro-x first flight, and~\cite{Manzagol:2021} for details on progress towards the 2022 reflight. The datasets generated during and/or analyzed during the current study are available from the corresponding author on reasonable request.

\begin{acknowledgements}
This work was supported by NASA Grants 80NSSC20K0430 and 80NSSC21K1856. Part of this work was performed under the auspices of the U.S. Department of Energy by Lawrence Livermore National Laboratory under Contract DE-AC52-07NA27344.
\end{acknowledgements}

\pagebreak

\end{document}